\documentclass[11pt]{article} \usepackage{mystyle-new}
\usepackage{epsfig,amsmath} \usepackage{graphics}
\usepackage{cite,./mcite}
 
  {\begin{list}{}{\settowidth{\labelwidth}{#1}%
  \setlength{\leftmargin}{\labelwidth}%
  \addtolength{\leftmargin}{\labelsep}%
  \setlength{\itemsep}{0pt plus 1pt}
  \setlength{\parsep}{0pt plus 1pt}
  \setlength{\topsep}{0pt plus 1pt}
  \setlength{\partopsep}{0pt plus 1pt}
  \setlength{\parskip}{2mm plus 1mm minus 1mm}
  }}%
  {\end{list}}
\def\lsim{\mathrel{\rlap{\lower4pt\hbox{\hskip1pt$\sim$}}
    \raise1pt\hbox{$<$}}}                
\def\gsim{\mathrel{\rlap{\lower4pt\hbox{\hskip1pt$\sim$}}
    \raise1pt\hbox{$>$}}}                

\newcommand{\cA}{{\cal A}}
\newcommand{\as}{\ensuremath{\alpha_\mathrm{s}}}

\def\prp{t}

\def\kt{\ensuremath{k_{\prp}}}
\def\shat{\ensuremath{\hat{s}}}

\newcommand{\alphasb}{\bar{\alpha}_s}

\def\cascade{{\sc Cascade3}}

\def\PYTHIA{{\sc Pythia}}

\def\tmdlib{{\sc TMDlib}}
\def\katie{{\sc KaTie}}
\def\powheg{{\sc Powheg}}

\def\herwig{{\sc Herwig}}
\def\pythia{{\sc Pythia}}
\def\mcatnlo{{\sc mc@nlo}}
\def\MGvATNLO{{\sc MadGraph5}\_a{\sc mc@nlo}}
\def\pegasus{{\sc Pegasus}}
\def\PB{{PB}}

\newenvironment{tolerant}[1]{\par\tolerance=#1\relax}{ \par }

\newcommand{\ccfm}{Ciafaloni:1987ur,Catani:1989yc,Catani:1989sg,Marchesini:1994wr}
\newcommand{\bfkl}{Kuraev:1976ge,Kuraev:1977fs,Balitsky:1978ic}

\usepackage{lineno}
\makeindex
\begin{document}
\begin{flushright}
DESY 21-005\\
EPJC published version
\end{flushright}
\begin{center} {\sffamily\Large\bfseries 
CASCADE3 \\ \vspace*{0.2cm}
A Monte Carlo event generator based on TMDs \\ \vspace*{0.2cm}
}
{ 
 \Large 
S.~Baranov$^1$, 
A.~Bermudez~Martinez$^{2}$, 
L.I.~Estevez~Banos$^{2}$,
F.~Guzman$^3$, 
F.~Hautmann$^{4,5}$,
H.~Jung$^{2}$,
A.~Lelek$^{4}$,
J.~Lidrych$^{2}$,
A.~Lipatov$^6$, 
M.~Malyshev$^6$,
M.~Mendizabal$^2$, 
S.~Taheri~Monfared$^{2}$,
A.M.~van~Kampen$^4$, 
Q.~Wang$^{2,7}$
H.~Yang$^{2,7}$
}\\  \vspace*{0.2cm}
     {\large $^1$Lebedev Physics Institute, Russia }\\
      {\large $^2$DESY, Hamburg, Germany}\\
     {\large $^3$InSTEC, Universidad de La Habana, Cuba}\\
    {\large $^4$Elementary Particle Physics, University of Antwerp, Belgium}\\
    {\large $^5$RAL and University of Oxford, UK} \\
    {\large $^6$SINP, Moscow State University, Russia }\\
    {\large $^7$School of Physics, Peking University, China} \\  

\end{center}

\begin{abstract}

The  \cascade\ Monte Carlo event generator based on Transverse Momentum Dependent (TMD) parton densities is described. 
Hard processes which are generated in collinear factorization with LO multileg or NLO parton level generators are extended by 
adding transverse momenta to the initial partons according to TMD densities and applying dedicated TMD parton showers and hadronization.
Processes with off-shell kinematics within \kt-factorization, either internally implemented  or from external packages via LHE files, can be processed for parton showering and hadronization.
The initial state parton shower is tied to the TMD parton distribution, with all parameters fixed by the TMD distribution.
\end{abstract} 


\section{Introduction}

The simulation of processes for high energy hadron colliders has been improved significantly in the past years by automation of next-to-leading order (NLO) calculations and matching of the hard processes to parton shower Monte Carlo event generators which also include a simulation of hadronization. Among those automated tools are the \MGvATNLO~\cite{Alwall:2014hca} generator based on the  \mcatnlo~\mcite{Frixione:2006gn,Frixione:2003ei,Frixione:2002bd,Frixione:2002ik}  method or the \powheg~\cite{Alioli:2010xa,Frixione:2007vw} generator for the calculation of the hard process. The results from these packages are then combined with either the \herwig~\cite{Bahr:2008pv} or \pythia~\cite{Sjostrand:2014zea} packages for parton showering and hadronization. Different jet multiplicities can be combined at the matrix element level and then merged with special procedures, like the MLM~\cite{Alwall:2007fs} or CKKW~\cite{Catani:2001cc} merging for LO processes,  the FxFx~\cite{Frederix:2012ps} or MiNLO method~\cite{Hamilton:2012np} for merging at NLO, among others.
While the approaches of matching and merging matrix element calculations and parton showers are very successful, two ingredients important for high energy collisions are not (fully) treated:  the matrix elements are calculated with collinear dynamics and the inclusion of initial state parton showers results in a net transverse momentum of the hard process;  the special treatment of high energy effects (small $x$) is not included. 

The {\sc Cascade} Monte Carlo event generator, developed originally for small $x$ processes based on high-energy factorization~ \cite{Catani:1990xk} and the CCFM~\cite{\ccfm} evolution equation, has been extended to cover the full kinematic range (not only small $x$) by applying the Parton Branching (\PB) method and the corresponding \PB\   Transverse Momentum Dependent (TMD) parton densities  \cite{Hautmann:2017fcj,Hautmann:2017xtx}. The initial state evolution is fully described and determined by the TMD density, as it was in the case of the CCFM gluon density, but now available for all flavor species, including quarks, gluons and photons at small and large $x$ and any scale $\mu$. For a general overview of TMD parton densities, see Ref.~\cite{Angeles-Martinez:2015sea}.

With the advances in determination of \PB\  TMDs  \cite{Hautmann:2017fcj,Hautmann:2017xtx}, it is natural to develop a scheme, where the initial parton shower follows as close as possible the TMD parton density and where either collinear (on-shell) or \kt-dependent (off-shell) hard process calculations can be included at LO or NLO.  In order to be flexible and to use the latest developments in automated matrix element calculations of hard process at higher order in the strong coupling $\alpha_s$, events available in the Les Houches Event (LHE) file format~\cite{Alwall:2006yp}, which contains all the information of the hard process including the color structure, can be further processed for parton shower and hadronization in  \cascade .

In this report we describe the new developments in \cascade\ for a full \PB -TMD parton shower and the matching of TMD parton densities to collinear hard process calculations. We also mention features of the small-$x$ mode of \cascade.

\section{The hard process}
\label{sec:HardProcess}
The cross section for the scattering process  of two hadrons $A$ and $B$ can be written in collinear factorization as a convolution of the partonic cross section of partons $a$ and $b$, $ a + b \to X$, and the densities $f_{a(b)}(x,\mu)$ of partons $a$ ($b$) inside the hadrons $A$ ($B$), 
\begin{equation}
\sigma(A+B\to Y ) =\int dx_a   \int dx_b   \,f_a(x_a,\mu) \,f_b(x_b,\mu) \,\sigma(a+b\to X) \, ,
\label{coll_x_section}
\end{equation}
where $x_a (x_b)$ are the fractions of the longitudinal momenta of hadrons  $A,B$  carried by the partons $a (b)$, $\sigma(a+b\to X)$ is the partonic cross section, and $\mu$ is the factorization scale of the process. The final state $Y$ contains the partonic final state $X$ and the recoils from the parton evolution and hadron remnants.

In \cascade\ we extend collinear factorization to include transverse momenta in the initial state, either by adding a transverse momentum to an on-shell process or by using off-shell processes directly, as described in detail in Sections~\ref{sec:OnShell} and~\ref{sec:OffShell}.
TMD factorization is proven for semi-inclusive deep-inelastic scattering, Drell-Yan production in hadron-hadron collisions and $e^+e^-$ 
annihilation~\cite{Collins:1981uk,Collins:1981uw,Collins:1982wa,Collins:1981tt,Collins:1984kg,Collins:2011zzd,Meng:1995yn,Nadolsky:1999kb,Nadolsky:2000ky,Ji:2004wu,Ji:2004xq,GarciaEchevarria:2011rb,Chiu:2011qc}. In the high-energy limit (small-$x$) $k_T$-factorization has been formulated also in hadronic collisions for processes like heavy flavor or heavy boson (including Higgs) production~\cite{Catani:1990xk,Levin:1991ry,Collins:1991ty,
Hautmann:2002tu}, with so-called {\em unintegrated} parton distribution functions (uPDFs), 
see {\it e.g.} Refs.~\cite{Avsar:2012hj,Avsar:2011tz,Jadach:2009gm,Dominguez:2011saa,Dominguez:2011br,Dominguez:2011gc,Hautmann:2009zzb,Hautmann:2012pf,Hautmann:2007gw,Catani:1993ww,Catani:1994sq}.

\subsection{On-shell processes}
\label{sec:OnShell}
\begin{tolerant}{3000}
The hard processes in collinear factorization (with on-shell initial partons, without transverse momenta) can be calculated by standard automated methods like 
 \MGvATNLO~\cite{Alwall:2014hca} 
for multileg processes at LO or  NLO accuracy.  The matrix element processes are calculated with collinear parton densities (PDF), as provided by LHAPDF~\cite{Buckley:2014ana}. 

We extend the factorization formula given in eq.(\ref{coll_x_section}) by replacing the collinear parton densities $f(x,\mu)$ by TMD densities ${\cal A}(x,\kt,\mu)$ with $\kt$ being the transverse momentum of the interacting parton, and  integrating over the transverse momenta.

However, when the hard process is to be combined with a TMD parton density, as described later, the integral over \kt\ of the TMD density must agree with the collinear (\kt -integrated) density; this feature is guaranteed by construction for the \PB-TMDs (also available as integrated PDFs in LHAPDF format).
\end{tolerant}
In a LO partonic calculation the TMD or the parton shower can be included respecting energy momentum conservation, as described below. 
In an NLO calculation based on the MC@NLO method ~\cite{Frixione:2006gn,Frixione:2003ei,Frixione:2002bd,Frixione:2002ik} the contribution from collinear and soft partons is subtracted, as this is added later with the parton shower. For the use with \PB\ TMDs, the \herwig 6 subtraction terms are best suited as the angular ordering conditions coincide with those applied in the \PB -method. The \PB\ TMDs play the same role as a parton shower does, in the sense that a finite transverse momentum is created as a result of the parton evolution~ \cite{Dooling:2012uw,Hautmann:2012dw}.

When transverse momenta of the initial partons from TMDs are to be included to the hard scattering process, which was originally calculated under the assumption of collinear initial partons, care has to be taken that energy and momentum are still conserved. When the initial state partons have transverse momenta, they also acquire virtual masses. 
The procedure adopted in \cascade\ is the following:
for each initial parton, a transverse momentum is assigned according to the TMD density, and the parton-parton system is boosted to its center-of-mass frame and rotated such that only the longitudinal and energy components are non-zero. The energy and longitudinal component of the initial momenta $p_{a,b}$ are recalculated taking into account the virtual masses $Q_a^2=k_{t,a}^2$ and $Q_b^2=k_{t,b}^2$  \cite{Bengtsson:1986gz}, 
\begin{eqnarray}
E_{a,b} &=& \frac{1}{2\sqrt{\shat}} \left( \shat \pm (Q_b^2 - Q_a^2) \right)\\
p_{z\;a,b} & = &  \pm \frac{1}{2\sqrt{\shat}}\sqrt{(\shat + Q_a^2 +Q_b^2)^2 - 4Q_a^2Q_b^2 }
\end{eqnarray}
with $\shat=(p_a + p_b)^2$ with $p_a (p_b)$ being the four-momenta of the interacting partons $a$ and $b$. The partonic system is then rotated and boosted back to the overall center-of-mass system of the colliding particles. By this procedure, the parton-parton mass $\sqrt{\hat{s}}$ is exactly conserved, while the rapidity of the partonic system is approximately restored, depending on the transverse momenta.

\begin{figure}[htb]
\begin{center} 
\includegraphics[width=0.325\textwidth]{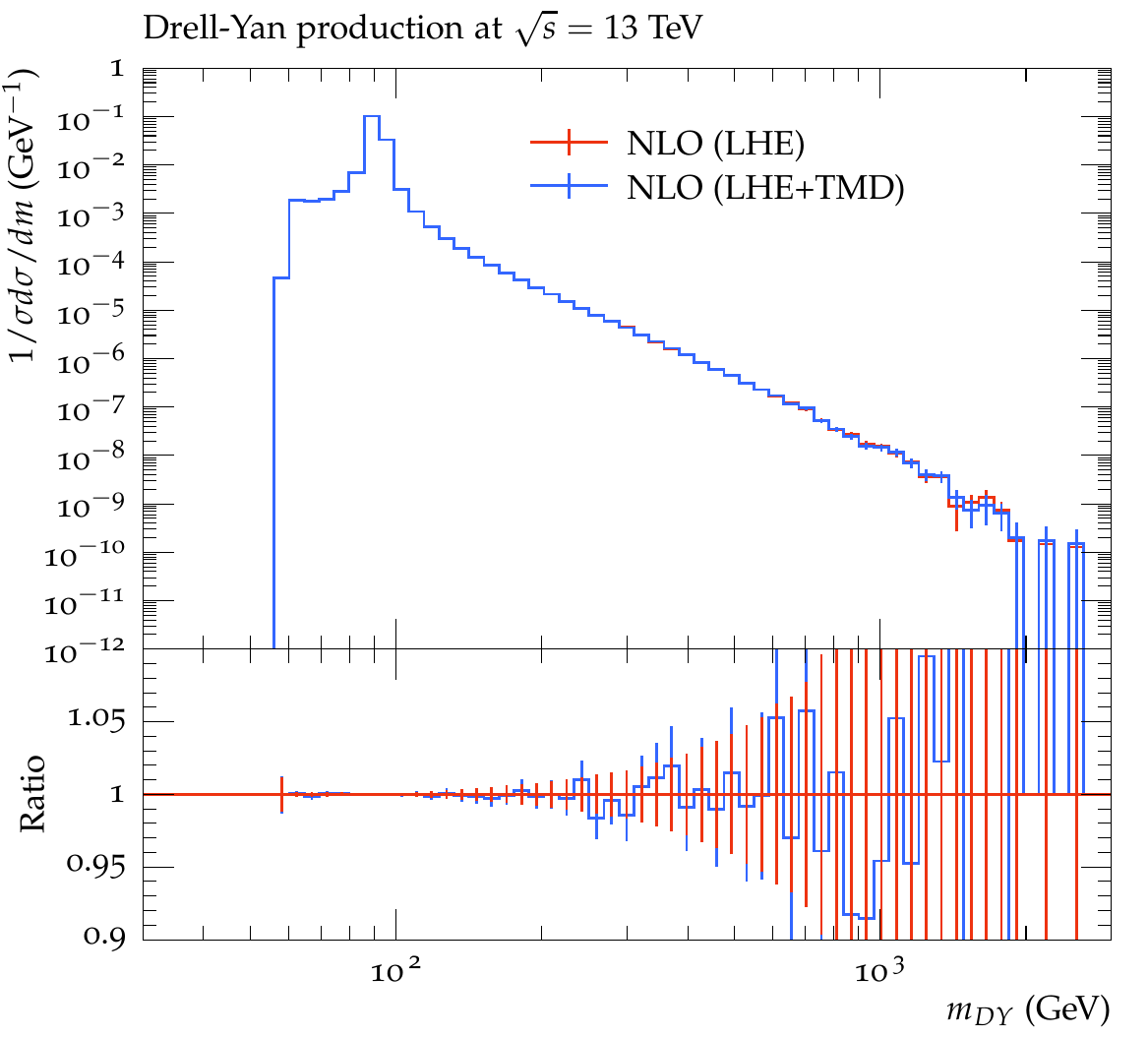}
\includegraphics[width=0.325\textwidth]{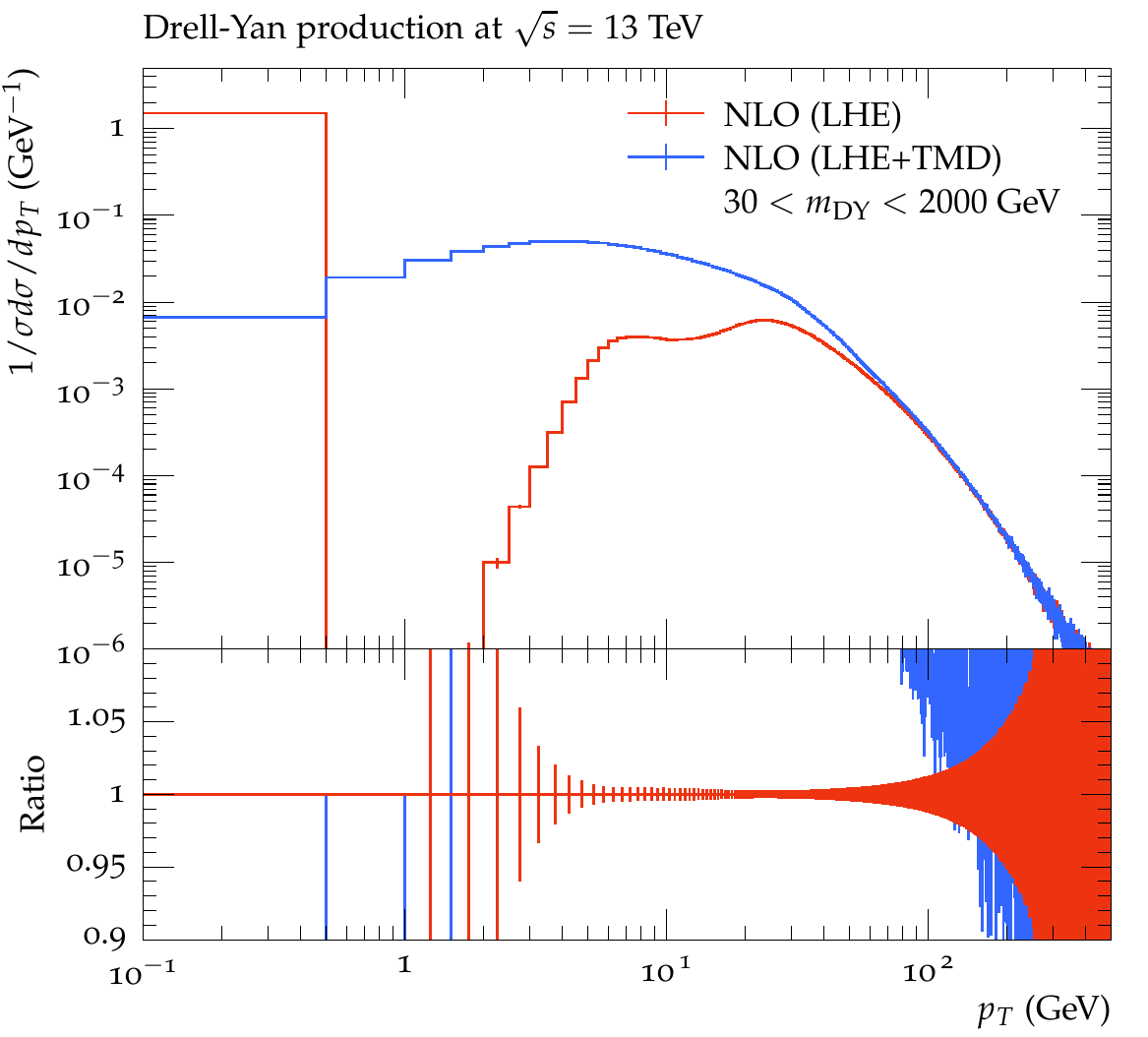}
\includegraphics[width=0.325\textwidth]{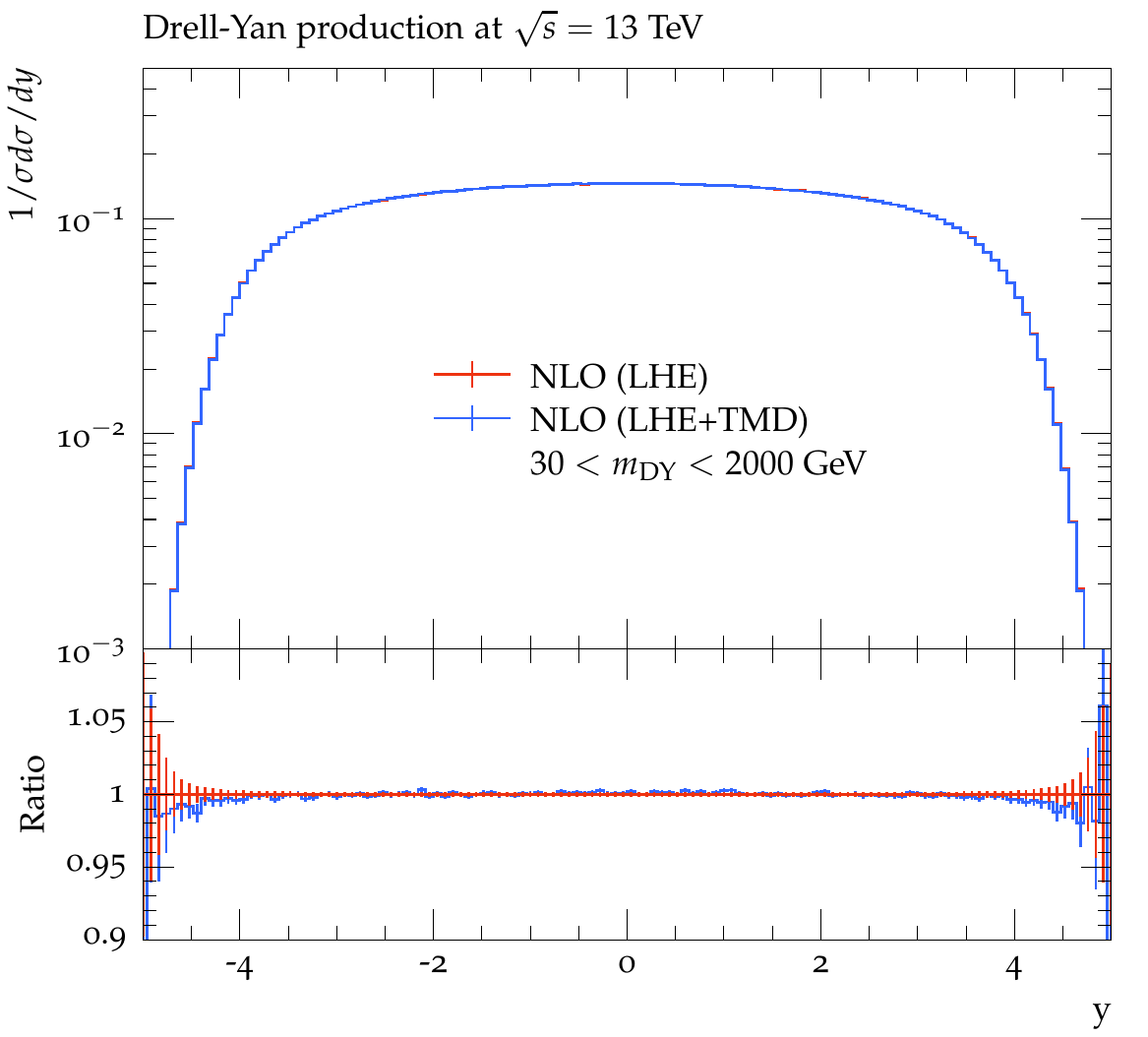}
\caption{Distributions of Drell-Yan mass, transverse momentum and rapidity for $pp \to DY + X$ at $\sqrt{s}=13 $ TeV. The hard process is calculated with \protect\MGvATNLO . NLO(LHE) is the prediction including subtraction terms, NLO(LHE+TMD) includes transverse momenta of the interacting partons according to the description in the text.}
\label{addTMD}
\end{center}
\end{figure}

In Fig.~\ref{addTMD} a comparison of the Drell-Yan (DY) mass, transverse momentum and rapidity is shown for an NLO calculation of DY production in pp collisions at $\sqrt{s}=13$ TeV in the mass range $30 < m_{DY} < 2000$ GeV. The curve labelled NLO(LHE) is the calculation of  \MGvATNLO\ with the subtraction terms, the curve NLO(LHE+TMD) is the prediction after the transverse momentum is included according to the procedure described above. 
In the $p_T$ spectrum one can clearly see the effect of including transverse momenta from the TMD distribution.
The DY mass distribution is not changed, and the rapidity distribution is almost exactly reproduced, only at large rapidities small differences are observed.

The transverse momenta $k_t$ are generated according to the TMD density ${\cal A}(x,k_{t},\mu)$, at the original longitudinal momentum fraction $x$ and the hard process scale $\mu$.
In a LO calculation, the full range of $\kt$ is available, but in an NLO calculation via  the MC@NLO method a {\it shower scale} defines the boundary between parton shower and real emissions from the matrix element, limiting the transverse momentum \kt .
Technically the factorization scale $\mu$ is calculated within \cascade\ (see parameter \verb+lhescale+)  as it is not directly accessible from the LHE file, while the {\it shower scale} is given by \verb+SCALUP+. 
The limitation of the transverse momenta coming from the TMD distribution and TMD shower to be smaller than the {\it shower scale} SCALUP guarantees that the overlap 
with real emissions from the matrix element is minimized according to the subtraction of counterterms in the MC@NLO method.

The advantage of using TMDs for the complete process is that the kinematics are fixed, independent of simulating explicitly the radiation history from the parton shower. For inclusive processes, for example inclusive Drell-Yan processes, the details of the hadronic final state generated by a parton shower do not matter, and  only the net effect of the transverse momentum distribution is essential. However, for processes which involve jets, the details of the parton shower become also important.
The parton shower, as described below, follows very closely the transverse momentum distribution of the TMD and thus does not change any kinematic distribution after the transverse momentum of the initial partons are included.

All hard processes, which are available in  \MGvATNLO\ 
can be used within \cascade . The treatment of multijet merging is described in Section~\ref{mulitjetMerging}.

\subsection{Off-shell processes}
\label{sec:OffShell}
In a region of phase space, where the longitudinal momentum fractions $x$ become very small, the transverse momentum of the partons cannot be neglected and has to be included already at the matrix element level, leading to so-called {\it off-shell} processes.

\begin{figure}[htb]
\begin{center} 
\includegraphics[width=0.49\textwidth]{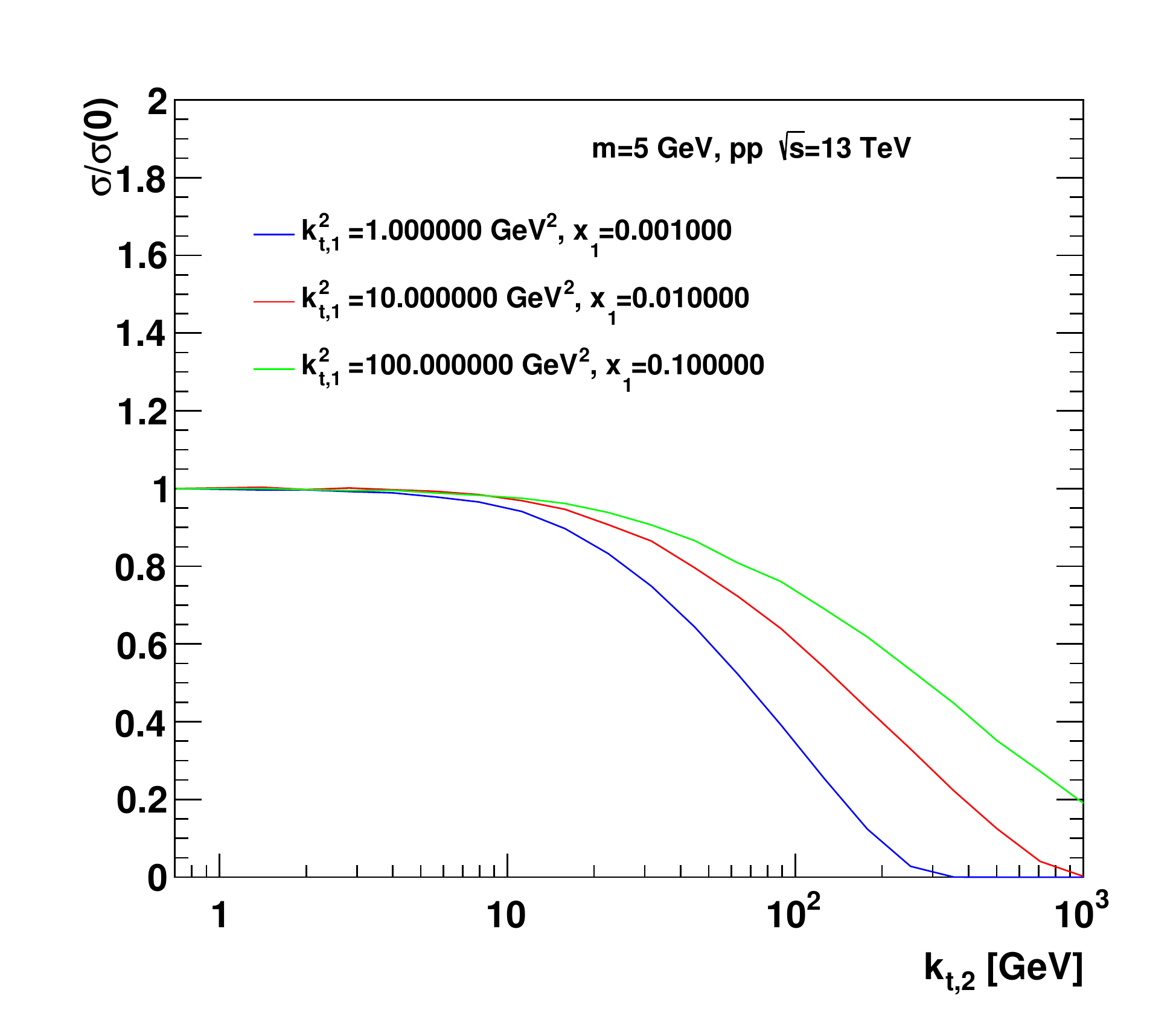}
\includegraphics[width=0.49\textwidth]{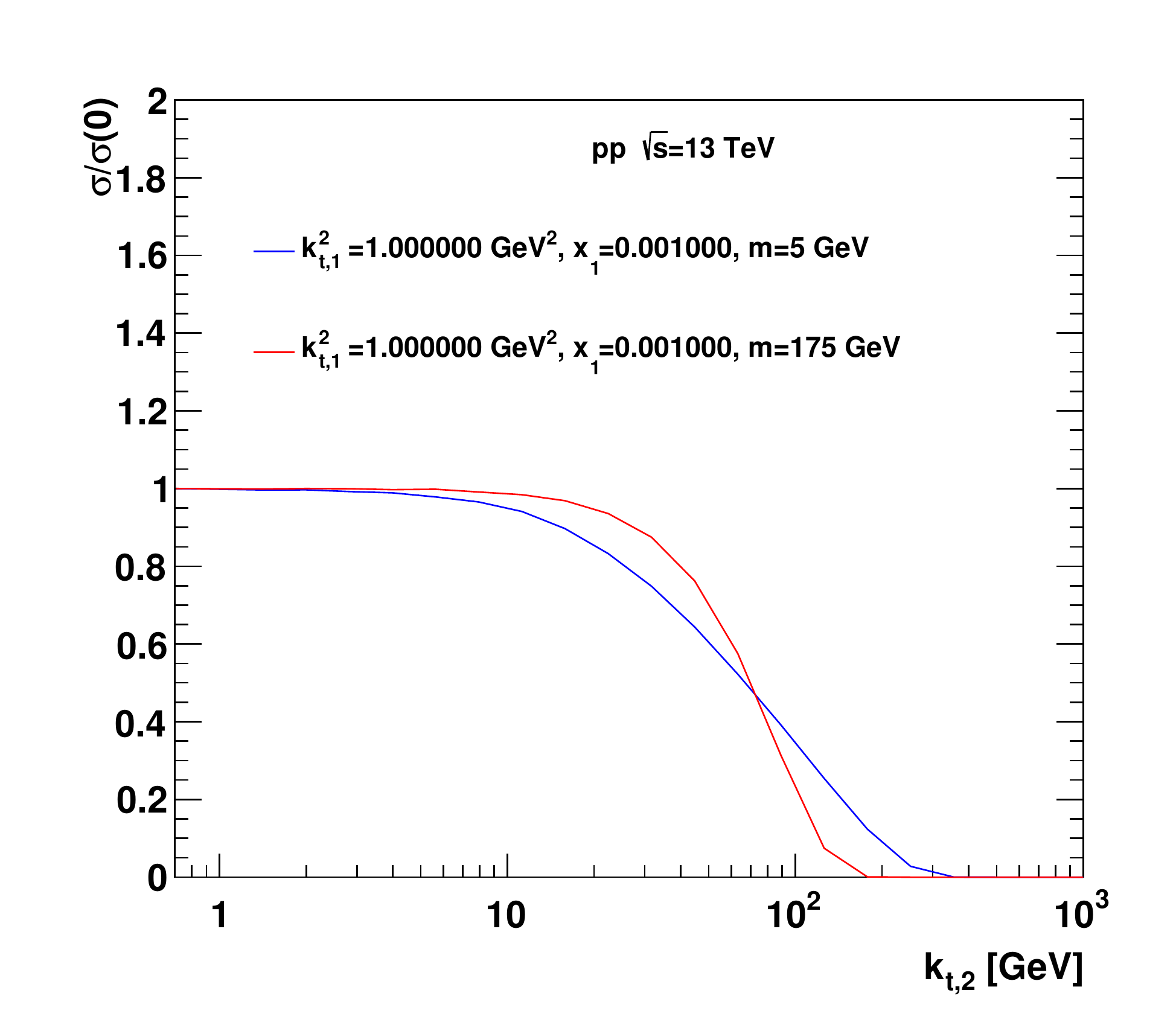}
\caption{The reduced cross section $\tilde{\sigma}(\kt)/ \tilde{\sigma}(0)$  as a function of the transverse momentum $k_{t,2}$ of the incoming gluon at $\sqrt{s} = 13000$ GeV. (Left) for different values of $k_{t,1}$ and $x_1$, (right) for different heavy flavor masses and
fixed values of $k_{t,1}$ and $x_1$. }
\label{offshell}
\end{center}
\end{figure}

In off-shell processes a natural suppression at large \kt\ \cite{Catani:1992rn} (with $\kt > \mu$)  is obtained, shown explicitly in Fig.~\ref{offshell}, where the matrix element for $g^* g^* \to Q\bar{Q}$, with $Q$ being a heavy quark, is considered. The process is integrated over the final state phase space~\cite{Marchesini:1992jw}, 
\begin{equation}
\tilde{\sigma}(\kt) = \int \frac{d x_2}{x_2}\, d \phi_{1,2}\,  {\rm d Lips}\, |ME|^2 \, (1-x_2)^5   \; , 
\end{equation} 
where ${\rm d Lips}$ is the Lorentz-invariant phase space of the final state, ${\rm ME}$ is the matrix-element for the process,  
$\phi_{1,2}$ is the azimuthal angle between the two initial partons, and a simple scale-independent and \kt -independent gluon density $xG(x)=(1-x)^5$ is included which suppresses 
large-$x$ contributions.  In Fig.~\ref{offshell} we show $\tilde{\sigma}(\kt)$ normalized to its on-shell value $\tilde{\sigma}(0)$  at $\sqrt{s}=13000$ GeV as a function of the transverse momentum of the incoming gluon $k_{t,2}$ 
for different values of $x_1$, which are chosen such that the ratio $k^2_{t,1}/ (x_1 s)$ is kept constant.

In Fig.~\ref{offshell} (left) predictions are shown for bottom quarks with mass $m=5$~GeV and different $k_{t,1}$, in Fig.~\ref{offshell} (right) 
a comparison is made for different heavy quark masses. Using off-shell matrix elements a suppression at large transverse momenta of the initial partons is obtained, depending on the heavy flavor mass and the transverse momentum. In a collinear approach, with implicit integration over transverse momenta of the initial state partons, the transverse momenta are limited by a theta function at the factorization scale, while off-shell matrix elements give a smooth transition to a high $k_t$ tail.

When using off-shell processes, BFKL or CCFM type parton densities should be used  
to cover the full available phase space in transverse momentum, which can lead to \kt 's 
 larger than the transverse momentum of any of the partons of the hard process~\cite{Hautmann:2014uua}. 
Until now, only gluon densities obtained from CCFM\cite{\ccfm} or BFKL\cite{\bfkl} are available, thus limiting the advantages of using off-shell matrix elements to gluon induced processes.

Several processes with off-shell matrix elements are implemented in \cascade\, as listed in Tab.~\ref{processes}, and described in detail in \cite{Jung:2010si}. However, many more processes are accessible via the automated matrix element calculators for off-shell processes, \katie ~\cite{vanHameren:2016kkz} and  \pegasus~\cite{Lipatov:2019oxs}. The events from the hard process are then read with the \cascade\ package via LHE files.
For processes generated with \katie\ or \pegasus\ no further corrections need to be performed and the event can be directly passed to the showering procedure, described in the next section.

\begin{table}[htp]
\begin{center}
\begin{tabular}{|l||l|c|c|}
\hline 
Lepto(photo)production &process & IPRO & Reference \\ 
\hline
&$\gamma^* g^* \to q\bar{q} $ & 10 & \protect\cite{Catani:1990eg} \\
&$\gamma^* g^* \to Q\bar{Q} $ & 11 & \protect\cite{Catani:1990eg} \\ 
&$\gamma^* g^* \to J/\psi g $ & 2 & \protect\cite{Saleev:1994fg,Lipatov:2002tc, Baranov:2003at,Baranov:2002cf} \\
\hline
Hadroproduction& & &  \\
\hline
&$g^* g^* \to q\bar{q} $ & 10 & \protect\cite{Catani:1990eg} \\
&$g^* g^* \to Q\bar{Q} $ & 11 & \protect\cite{Catani:1990eg} \\ 
&$g^* g^* \to J/\psi g $ & 2 & \protect\cite{Baranov:2002cf} \\
&$g^* g^* \to \Upsilon g $ & 2 & \protect\cite{Baranov:2002cf} \\
&$g^* g^* \to \chi_c $ & 3 & \protect\cite{Baranov:2002cf} \\
&$g^* g^* \to \chi_b $ & 3 &\cite{Baranov:2002cf} \\
&$g^* g^* \to J/\psi J/\psi $ & 21 &\cite{Baranov:2011zz} \\
&$g^* g^* \to h^0 $ & 102 & \protect\cite{Hautmann:2002tu} \\
&$g^* g^* \to Z Q \bar{Q} $ & 504 & \protect\cite{Baranov:2008hj,Deak:2008ky} \\
&$g^* g^* \to Z q \bar{q} $ & 503 & \protect\cite{Baranov:2008hj,Deak:2008ky} \\
&$g^* g^* \to W q_i Q_j $ & 514 & \protect\cite{Baranov:2008hj,Deak:2008ky} \\
&$g^* g^* \to W q_i q_j $ & 513 & \protect\cite{Baranov:2008hj,Deak:2008ky} \\
&$q g^* \to Z q $ & 501 & \protect\cite{Marzani:2008uh} \\
&$q g^* \to W q $ & 511 & \protect\cite{Marzani:2008uh} \\
&$q g^* \to qg $ & 10 & \protect\cite{Deak:2009xt} \\
&$g g^* \to gg $ & 10 & \protect\cite{Deak:2009xt}\\
\hline
\end{tabular}
\caption{\it Processes included in \protect\cascade . $Q$ stands for heavy quarks, $q$ for light quarks.}\label{processes}
\end{center}
\end{table}%

\section{Initial State Parton Shower based on TMDs}

The parton shower, which is described here, follows consistently the parton evolution of the TMDs. 
By this we mean that the splitting functions $P_{ab}$, the order and  the scale in \as\, as well as  kinematic restrictions are identical to both the parton shower and the evolution of the parton densities (for NLO \PB\ TMD densities, the NLO DGLAP splitting functions~\cite{Furmanski:1981cw,Curci:1980uw} together with NLO $\as$ is applied, while for the LO TMD densities the corresponding LO splitting functions ~\cite{Dokshitzer:1977sg,Altarelli:1977zs,Gribov:1972ri} and LO $\as$ is used).

\subsection{From \PB\ TMD evolution to TMD Parton Shower}

The \PB\ method describes the TMD parton density as (cf eq.(2.43) in Ref. \cite{Hautmann:2017fcj}) 
\begin{eqnarray}
\label{integeqforA}
  { x{\cal A}}_a(x,k_t, \mu) &=&  \Delta_a (  \mu  ) \  x { {\cal A}}_a(x,k_t,\mu_0)  
 + \sum_b \int {{d q^{ 2} } \over {q^{ 2} } }  {{d \phi  } \over {2 \pi  }}  \  
{
{\Delta_a (  \mu  )} 
 \over 
{\Delta_a (  q  
 ) }
}
\ \Theta(\mu-q) \  
\Theta(q - \mu_0)
 \nonumber\\ 
&\times&  
\int_x^{z_M} {dz} \;
P_{ab}^{(R)} (\as(f(z,q))
,z) 
\;\frac{x}{z} { {\cal A}}_b\left({\frac{x}{z}}, k_t^{\prime} , 
q\right)  
  \;\;  ,     
\end{eqnarray}
with  $z_M<1$ defining resolvable branchings, ${\bf k}$ (${\bf q}_c$) being the transverse momentum vector of the propagating (emitted) parton, respectively.
 The transverse momentum of the parton before branching is defined as 
$k_t^{\prime } = | {\bf k}+(1-z) {\bf q}| $ with  ${\bf q} = {\bf q}_c/(1-z) $ being the rescaled transverse momentum vector of the emitted parton (see Fig.~\ref{parton-branching}, with the notation  $k_t=|{\bf k}|$ and $q=|{\bf q}|$) and $\phi$ being the azimuthal angle between $ {\bf q}$ and ${\bf k}$.
The argument in $\as$ is in general a function of the evolution scale $q$. Higher order calculations indicate the transverse momentum of the emitted parton as the preferred scale. The real emission branching probability is denoted by $P_{ab}^{(R)} (\as(f(z,q)), z)$ including $\as$ as described in Ref.~\cite{Hautmann:2017fcj} (in the following we omit $\as$   in the argument of $ P_{ab}^{(R)}$ for easier reading). 
The Sudakov form factor is given by:  
\begin{equation}
\label{sud-def}
 \Delta_a ( z_M, \mu , \mu_0 ) = 
\exp \left(  -  \sum_b  
\int^{\mu^2}_{\mu^2_0} 
{{dq^{ 2} } 
\over q^{ 2} } 
 \int_0^{z_M} dz \  z 
\ P_{ba}^{(R)} 
\right) 
  \;.
\end{equation}
Dividing Eq.(\ref{integeqforA}) by $\Delta_a (  \mu^2  )$ and differentiating with respect to  ${ \mu^{ 2 } }$  gives the differential form of the evolution equation describing
the probability for resolving a parton with transverse momentum ${\bf k}^\prime$ and momentum fraction $x/z$ into a parton with momentum fraction $x$ and emitting another parton
 during a small decrease of $\mu$, 
\begin{eqnarray}
\label{diffeqforA}
 { \mu^2} \frac{d} {d \mu^2 }
\left( \frac{ { x{\cal A}}_a(x, k_t, \mu) }{ \Delta_a (  \mu )} \right) & =  &
 \sum_b 
\int_x^{z_M} {dz} {{d \phi  } \over {2 \pi  }} \;
P_{ab}^{(R)}
\;\frac{x}{z} \frac{{ {\cal A}}_b\left({\frac{x}{z}}, k_t^{\prime }, 
\mu \right)}  { \Delta_a ( \mu)}   \;. 
  \;\;      
\end{eqnarray}
The normalized probability is then given by 
\begin{eqnarray}
\label{normProb}
\frac{ \Delta_a ( \mu  )}{ { x{\cal A}}_a(x,k_t, \mu) } d\left( \frac{ { x{\cal A}}_a(x,k_t, \mu) }{ \Delta_a (  \mu  )} \right) &=& 
 \sum_b {{d \mu^2 } \over {\mu^2} } 
\int_x^{z_M} {dz} {{d \phi  } \over {2 \pi  }} \;
P_{ab}^{(R)}  
\; \frac{{\frac{x}{z} {\cal A}}_b\left({\frac{x}{z}}, k_t^{\prime } , 
\mu \right) }{{ x{\cal A}}_a(x,k_t, \mu)}
  \;\;      
\end{eqnarray}
This equation can be integrated between $\mu^2_{i-1}$ and $\mu^2$ to give the no-branching probability (Sudakov form factor) for the backward evolution $\Delta_{bw}$,\footnote{In  Eq.(\ref{Sudakov_bw}) ordering in $\mu$ is assumed. However, if  angular ordering   as in CCFM~\cite{\ccfm}   is applied  then the ratio of parton densities would change to $ [x^{\prime}\cA_b(x^{\prime},k_t^{\prime },q^{\prime}/z)] / [ x\cA_a(x,k_t,q^{\prime})]$ as discussed in \cite{Jung:2010si}.}
\begin{eqnarray}
\label{Sudakov_bw}
\log \Delta_{bw}(x,k_t,\mu,\mu_{i-1}) &=&
\log \left( 
\frac{ \Delta_a (  \mu)}{ \Delta_a (  \mu_{i-1}  )}   
\frac{  x{\cal A}_a(x,k_t, \mu_{i-1}) } { { { x{\cal A}}_a(x,k_t, \mu) }
} \right)  
\\ \nonumber
&= & 
- \sum_b \int_{\mu_{i-1}^2}^{\mu^2}  {{d q^{\prime\,2 } } \over {q^{\prime\,2}} } {{d \phi  } \over {2 \pi  }}
\int_x^{z_M} {dz} \;
P_{ab}^{(R)} \; \frac{{x^\prime {\cal A}}_b\left(x^\prime, k_t^{\prime } , 
q^{\prime }\right) }{{ x{\cal A}}_a(x, k_t, q^{\prime })}
  \;,     
\end{eqnarray}
with $x^\prime=x/z$.
This Sudakov form factor is very similar to the Sudakov form factor in ordinary parton shower approaches, with the difference that for the \PB\ TMD shower the 
ratio of  \PB\ TMD densities  $ [x^\prime {\cal A}_b\left(x^\prime, k_t^{\prime } ,q^{\prime }\right) ]/ [ x{\cal A}_a(x, k_t, q^{\prime }) ]$ is applied, which includes a dependence on $k_t$.

In Eq.(\ref{Sudakov_bw}) a relation between the Sudakov form factor  $\Delta_a$ used in the evolution equation and the Sudakov form factor $\Delta_{bw}$ used for the backward evolution of the parton shower is made explicit. A similar relation was also studied in Refs.~\cite{Nagy:2020gjv,prestel:2020}. 
In Ref~\cite{Nagy:2020gjv} the $z_M$ limit was identified as a source of systematic uncertainty when using conventional showers with standard collinear pdfs; in the PB approach, the same $z_M$ limit is present in the parton evolution as well as in the PB-shower.
The \PB\ approach allows a consistent formulation of the parton shower with the \PB\ TMDs, as in both Sudakov form factors  $\Delta_a$ and $\Delta_{bw}$ the same value of $z_M$ is used.

The splitting functions $P_{ab}^{(R)} $ contain the coupling,  
\begin{equation}
P_{ab} (\as ,z) = \sum^{\infty}_{n=1} \left( \frac{\as(f(z,q)) }{2\pi}
\right)^{n} P_{ab}^{(n-1)}(z)\; ,
\label{Pab}
\end{equation}
where the scale $f( z,q)$ in the coupling depends on the ordering condition as discussed later (see Eq.(\ref{ang-ordering})).

The advantage of using a \PB\ TMD shower is that as long as the parameters of the parton shower are set through TMD distributions the parton shower uncertainties can be recast as uncertainties of the TMDs, which in turn can be fitted to experimental data in a systematic global manner.

\subsection{Backward Evolution for initial state TMD Parton Shower}
\label{sec:TMDshower}

A backward evolution method, as now common in Monte Carlo event generators, is applied for the initial state parton shower, evolving from the large scale of the matrix-element process backwards down to the scale of the incoming hadron. However, in contrast to the conventional parton shower, which generates transverse momenta of the initial state partons during the backward evolution, the transverse momenta of the initial partons of the hard scattering process is fixed by the TMD and the parton shower does not change the kinematics.
The transverse momenta during the backward cascade follow the behavior of the TMD.  The hard scattering process is obtained as described in section~\ref{sec:HardProcess}. 
The backward evolution of the initial state parton shower follows very closely the description in \cite{Jung:2010si,Jung:2001hx,Jung:2000hk}, which is based on Ref.~\cite{Bengtsson:1986gz}. 

The starting value of the evolution scale $\mu$ is calculated from the hard scattering process, as described in Section~\ref{sec:HardProcess}. In case of on-shell matrix elements at NLO, the transverse momentum of the hardest parton in the parton shower evolution is limited by the {\it shower-scale}, as described in Section~\ref{sec:OnShell}.

\begin{figure}[htb]
\begin{center} 
\includegraphics[width=0.25\textwidth]{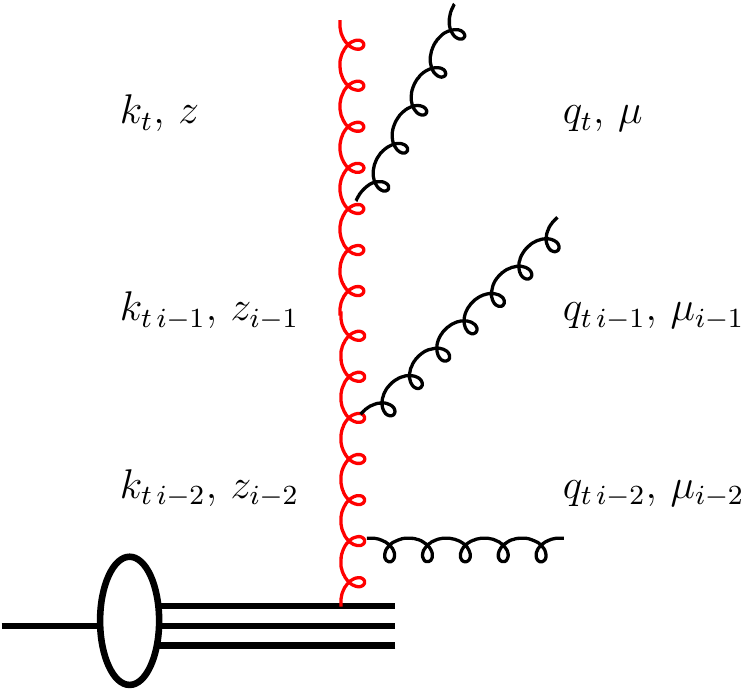} \hskip 2cm
\includegraphics[width=0.2\textwidth]{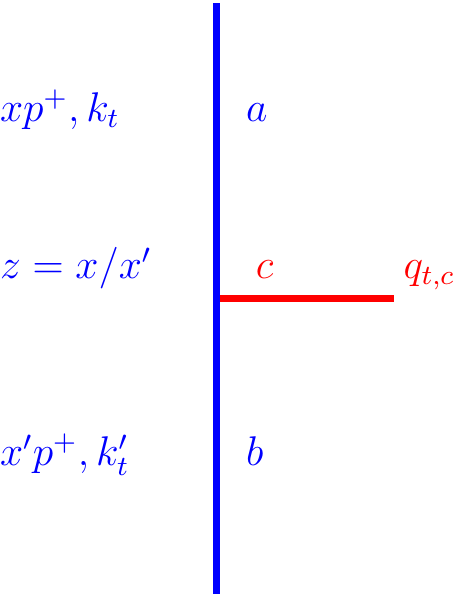} 
  \caption{Left: Schematic view of a parton branching process. Right: Branching process $ b \to a + c$.}
\label{parton-branching}
\end{center}
\end{figure}

Starting at the hard scale $\mu=\mu_{i}$, the  parton shower algorithm searches for the next scale $\mu_{i-1}$ at which a resolvable branching occurs (see Fig.~\ref{parton-branching} left). This scale $\mu_{i-1}$ is selected from the Sudakov form factor $\Delta_{bw}$ as given in Eq.(\ref{Sudakov_bw})  (see also \cite{Jung:2010si}). 
In the parton shower language, the selection of the next branching comes from solving  $R=\Delta_{bw}(x,k_t,\mu_{i},\mu_{i-1})$ for $\mu_{i-1}$  using uniformly distributed random numbers R for given $x$ and $\mu_{i}$.
However, to solve the integrals in Eq.(\ref{Sudakov_bw}) numerically for every branching would be too time consuming, instead the veto-algorithm \cite{Bengtsson:1986gz,Platzer:2011dq} is applied. 

The splitting function $P_{ab}$ as well as the argument $f(z,q)$ in the calculation of \as\ is chosen exactly as used in the evolution of the parton density. In a parton shower one treats ``resolvable'' branchings, defined via a cut in $z < z_M$ in the splitting function  to avoid the singular behavior of the terms $1/(1-z)$, and branchings with $z> z_M$ are regarded as ``non-resolvable'' and are treated similarly as virtual corrections: they are included in the Sudakov form factor $\Delta_{bw} $. 
The  splitting variable $z_{i-1}$ is obtained from the splitting functions following the standard methods (see Eq.(2.37) in \cite{Hautmann:2017fcj}). 

The calculation of the  transverse momentum $\kt$ is sketched in Fig.~\ref{parton-branching} (right).
The transverse momentum $q_{t\,c}$ 
can be calculated in case of angular ordering (where the scale $q$ of each branching is associated with the angle of the emission)  in terms of the angle $\Theta$ of the emitted parton with respect to the beam directions $q_{t,c} = (1-z) E_{b} \sin \Theta$, 
\begin{equation}
  \label{ang-ordering}
 {\bf q}_{c}^2  =  (1-z)^2 q^{ 2}  \;\; .
\end{equation}

Once the transverse momentum of the emitted parton ${\bf q}_c$ is known, the transverse momentum of the propagating parton can be calculated from
\begin{equation}
{\bf k}^\prime = {\bf k} + {\bf q}_{c}
\label{kt-definition}
\end{equation}
with a uniformly distributed azimuthal angle $\phi$ assumed for the vector components of ${\bf k}$ and ${\bf q}_c$. The generation of the parton momenta is performed in the center-of-mass frame of the collision (in contrast to conventional parton showers, which are generated in different partonic frames).

The whole procedure is iterated until one reaches a scale $\mu_{i-1} < q_0$ with $q_0$ being a cut-off parameter, which can be chosen to be the starting evolution scale of the TMD. It is of advantage to continue the parton shower evolution to lower  scales $q_0 \sim \Lambda_{qcd} \sim 0.3$~GeV.

The final transverse momentum of the propagating parton 
${\bf k }$   is the sum of all transverse momenta ${\bf q}_c$ (see Fig.~\ref{parton-branching} right):
\begin{equation}
 {\bf k } =  {\bf k }_0 - \sum_c {\bf q}_c    \;\; .
\label{kt_calc}
\end{equation}
with  ${\bf k }_0 $ being the intrinsic transverse momentum.

The \PB\ TMD parton shower is selected with \verb+PartonEvolution=2+ (or \verb+ICCF=2+).

\subsection{CCFM parton evolution and parton shower} 
\label{sec:CCFMshower}
The CCFM parton evolution and corresponding parton shower follows a similar approach as described in the previous section and in detail also in Refs.~\cite{Jung:2000hk,Jung:2001hx,Jung:2010si,Hautmann:2008vd}. The main difference to the \PB -TMD shower are the splitting functions with the non-Sudakov form factor $\Delta_{ns}$ and the allowed phase space for emission.
The original CCFM splitting function $\tilde{P}_g (z,q,\kt)$ for branching $g\to g g$  is given by\footnote{Finite terms are neglected as they are not obtained in CCFM at the leading infrared accuracy~(cf p.72 in \cite{Catani:1989sg}).} 
\begin{equation}
\tilde{P}_g (z,q,\kt)
= \frac{\alphasb(q(1-z))}{1-z} + 
\frac{\alphasb(\kt)}{z} \Delta_{ns}(z,q,\kt),
\label{Pgg}
\end{equation}
where the non-Sudakov form factor $\Delta_{ns}$ is defined as 
\begin{equation}
\log\Delta_{ns} =  -\alphasb(\kt)
                  \int_0^1 \frac{dz'}{z'} 
                        \int \frac{d q^2}{q^2} 
              \Theta(\kt-q)\Theta(q-z'q_{t}) \, ,
                  \label{non_sudakov}                   
\end{equation}
with $q_t= \sqrt{{\bf q}_{t}^2}$ being the magnitude of the transverse vector defined in Eq.(\ref{ang-ordering}) and \kt\  the magnitude of the 
transverse vector   in Eq.(\ref{kt-definition}).

The CCFM  parton shower is selected with \verb+ICCF=1+ ( \verb+PartonEvolution=1+). \footnote{A one loop parton shower (DGLAP like) with $\Delta_{ns}=1$, one loop $\as$ and strict ordering in $q$ 
can be selected with \texttt{ICCF=0}.}

\section{The TMD parton densities}

In the previous versions of {\sc Cascade}  the TMD densities were part of the program. With the development of \tmdlib ~\cite{Hautmann:2014kza,Abdulov:2021ivr}  there is easy access to all available TMDs, including parton densities for photons (as well as Z, W and H densities, if available).
 \begin{tolerant}{2000}
These parton densities  can be selected via \verb+PartonDensity+ with a value $>100000$.
For example the TMDs from the parton branching method \cite{Hautmann:2017fcj,Hautmann:2017xtx} are selected via  \verb+PartonDensity=102100 (102200)+ for PB-NLO-HERAI+II-2018-set1 (set2). 
\end{tolerant}

Note that the features of the TMD parton shower are only fully available for the \PB -TMD sets and the CCFM shower clearly needs CCFM parton densities  (like for instance \cite{Hautmann:2013tba}). 
\PB -TMD parton densities are determined in Ref.~\cite{Martinez:2018jxt} from fits to HERA DIS $F_2$ measurements for $Q^2 > 3 $ GeV$^2$, giving very good $\chi^2$ values.
In Refs. \cite{Martinez:2020fzs,Martinez:2019mwt} the transverse momentum distribution of  Drell-Yan pairs at low and high masses, 
obtained from \PB -TMD densities, are compared with experimental measurements in a wide variety of kinematic regions, from low-energy fixed target experiments to high-energy collider experiments. Good agreement is found between predictions and measurements without the need for tuning of nonperturbative parameters, which illustrates the validity of the approach over a broad kinematic range in energy and mass scales.

\section{Final state parton showers}

The final state parton shower uses the parton shower routine \verb+PYSHOW+  of \PYTHIA .
Leptons in the final state, coming for example from Drell-Yan decays, can radiate photons, which are also treated in the final state parton shower. Here the method from \verb+PYADSH+ of \PYTHIA\ is applied, with the scale for the QED shower being fixed at the virtuality of the decaying particle (for example the mass of the Z-boson). 

\begin{tolerant}{4000}
The default scale for the QCD final state shower is $\mu^2=2\cdot (m_{1\;\perp}^2+m_{2\;\perp}^2)$ (\verb+ScaleTimeShower=1+),
with $m_{1(2)\;\perp}$ being the transverse mass of the hard parton 1(2). Other choices are possible: $\mu^2=\hat{s}$ (\verb+ScaleTimeShower=2+) and $\mu^2=2\cdot (m_1^2+m_2^2)$ (\verb+ScaleTimeShower=3+). 
In addition a scale factor can be applied: \verb+ScaleFactorFinalShower+$\times\mu^2 $ (default:  \verb+ScaleFactorFinalShower=1+).
\end{tolerant}

\section{Hadronization}
\begin{tolerant}{1000}
The hadronization (fragmentation of the partons in colorless systems) is done exclusively by \PYTHIA . 
Hadronization (fragmentation) is switched off by \verb+Hadronization = 0+ (or \verb+NFRA = 0+ for the older steering cards).
All parameters of the hadronization model can be changed via the steering cards.
\end{tolerant}

\section{Uncertainties}
Uncertainties of QCD calculations mainly arise from missing higher order corrections, which are estimated by varying the factorization and renormalization scales up and down by typically a factor of 2. The scale variations are performed when calculating the matrix elements and are stored as additional weights in the LHE file, which are then passed directly via \cascade\ to the HEPMC~\cite{Dobbs:2001ck} output file for further processing.

The uncertainties coming from the PDFs can also be calculated as additional weight factors during the matrix element calculation. However, when using TMDs, additional uncertainties arise from the transverse momentum distribution of the TMD. The \PB -TMDs come with uncertainties from the experimental uncertainties as well as from model uncertainties, as discussed in Ref.~\cite{Martinez:2018jxt}. These uncertainties can be treated and applied as additional weight factors with the parameter \verb+Uncertainty_TMD=1+.

\section{Multi-jet merging}
\label{mulitjetMerging}

Showered multijet LO matrix element calculations can be merged using the prescription discussed in Ref.~\cite{PB-MLM}. The merging performance is controlled by the three parameters \verb+Rclus+, \verb+Etclus+, \verb+Etaclmax+. Final-state partons  with pseudorapidity $\eta<$\verb+Etaclmax+ present in the event record after the shower step but before hadronization are passed to the merging machinery if \verb+Imerge = 1+. Partons are clustered using the kt-jet algorithm with a cone radius \verb+Rclus+ and matched to the PB evolved matrix element partons if the distance between the parton and the jet is $R< 1.5\times$\verb+Rclus+. The hardness of the reconstructed jets is controlled by its minimum transverse energy \verb+Etclus+ (merging scale).

\begin{tolerant}{3000}
The number of light flavor partons is defined by the \verb+NqmaxMerge+ parameter. Heavy flavor partons and their corresponding radiation are not passed to the merging algorithm. All jet multiplicities are treated in exclusive mode except for the highest multiplicity \verb+MaxJetsMerge+ which is treated in inclusive mode. 
\end{tolerant}

\section{Program description}
In \cascade\ all variables are declared as \verb"Double Precision".  With \cascade\ the source of  \PYTHIA\ 6.428 is included to avoid difficulties in linking.

\subsection{Random Numbers}
\cascade\ uses the \verb+RANLUX+ random number generator, with luxory level      \verb+LUX = 4+. The random number seed can be set via the environment variable \verb+CASEED+, the default value is \verb+CASEED=12345+.

\subsection{Event Output}

When \verb+HEPMC+ is included, generated events are written out in HEPMC ~\cite{Dobbs:2001ck}  format for further processing. The environment variable \verb+HEPMCOUT+ is used to specify the file name, by default this variable is set to \verb+HEPMCOUT=/dev/null+.

 The HEPMC events can be further processed, for example with Rivet \cite{Buckley:2010ar}. 

\subsection{Input parameters}

The input parameters are steered via steering files. The new format of steering is discussed in Section~\ref{sec:NewSteer} and should be used when reading LHE files, while the other format, which is appropriate for the internal off-shell processes, is discussed in Section~\ref{sec:OldSteer}.

\subsubsection{Input parameters - new format}
\label{sec:NewSteer}

Examples for steering files are under \verb+$install_path/share/cascade/LHE+.

\begin{tolerant}{3000}
\begin{footnotesize}
\begin{verbatim}
&CASCADE_input
NrEvents = -1                  ! Nr of events to process
Process_Id = -1                ! Read LHE file
Hadronisation = 0              ! Hadronisation (on =1, off = 0)
SpaceShower = 1                ! Space-like Parton Shower
SpaceShowerOrderAlphas=2       ! Order alphas in Space Shower 
TimeShower = 1                 ! Time-like Parton Shower
ScaleTimeShower = 4            ! Scale choice for Time-like Shower
!                                 1: 2(m^2_1t+m^2_2t)
!                                 2: shat
!                                 3: 2(m^2_1+m^2_2)
!                                 4: 2*scalup (from lhe file)
!ScaleFactorFinalShower = 1.   ! scale factor for Final State Parton Shower
PartonEvolution = 2            ! type of parton evolution in Space-like Shower
!                                 1: CCFM
!                                 2: full all flavor TMD evolution
! EnergyShareRemnant = 4       ! energy sharing in proton remnant
!                                 1: (a+1)(1-z)**a, <z>=1/(a+2)=1/3
!                                 2: (a+1)(1-z)**a, <z>=1/(a+2)=mq/(mq+mQ
!                                 3: N/(z(1-1/z-c/(1-z))**2), c=(mq/mQ)**2
!                                 4: PYZDIS: KFL1=1
! Remnant = 0                  ! =0 no remnant treatment
PartonDensity = 102200           ! use TMDlib: PB-TMDNLO-set2
! PartonDensity = 102100         ! use TMDlib: PB-TMDNLO-set1
! TMDDensityPath= './share'    ! Path to TMD density for internal files
Uncertainty_TMD = 0                        ! calculate and store uncertainty TMD pdfs
lheInput='MCatNLO-example.lhe' ! LHE input file
lheHasOnShellPartons = 1       ! = 0 LHE file has off-shell parton configuration
lheReweightTMD = 0             ! Reweight with new TMD  given in PartonDensity
lheScale = 2                   ! Scale defintion for TMD
!                                 0: use scalup
!                                 1: use shat
!                                 2: use 1/2 Sum pt^2 of final parton/particles
!                                 3: use shat for Born and 1/2 Sum pt^2 of final parton(particle)
!                                 4: use shat for Born and max pt of most forward/backward 
!                                    parton(particle)
lheNBornpart = 2               ! Nr of hard partons (particles) (Born process)
ScaleFactorMatchingScale = 2.  ! Scale factor for matching scale when including TMDs
&End


&PYTHIA6_input
P6_Itune = 370                 ! Retune of Perugia 2011 w CTEQ6L1 (Oct 2012)
! P6_MSTJ(41) = 1              ! (D = 2) type of branchings allowed in shower.
!                                 1: only QCD
!                                 2: QCD and photons off quarks and leptons
P6_MSTJ(45) = 4                ! Nr of flavors in final state shower: g->qqbar
P6_PMAS(4,1)= 1.6              ! charm mass
P6_PMAS(5,1)= 4.75             ! bottom mass
P6_MSTJ(48) =   1              ! (D=0), 0=no max. angle, 1=max angle def. in PARJ(85)
! P6_MSTU(111) =  1            ! = 0 : alpha_s is fixed, =1 first order; =2 2nd order;
! P6_PARU(112) = 0.2           ! lambda QCD
P6_MSTU(112)=   4              ! nr of flavours wrt lambda_QCD
P6_MSTU(113)=                  ! min nr of flavours for alphas
P6_MSTU(114)=   5              ! max nr of flavours for alphas
&End
\end{verbatim}
\end{footnotesize}
\end{tolerant}

\subsubsection{Input parameters - off-shell processes}
\label{sec:OldSteer}
\begin{tolerant}{3000}
Examples for steering files are under \verb+$install_path/share/cascade/HERA+ and \verb+$install_path/share/cascade/PP+.
\end{tolerant}

\begin{footnotesize}
\begin{verbatim}
* OLD STEERING FOR CASCADE 
*
* number of events to be generated
*
NEVENT 100
*
* +++++++++++++++++ Kinematic parameters +++++++++++++++
*
'PBE1'    1    0    -7000.    ! Beam energy                                 
'KBE1'    1    0     2212     ! -11: positron, 22: photon  2212: proton 
'IRE1'    1    0       1      ! 0: beam 1 has no structure 
*                             ! 1: beam 1 has structure  
'PBE2'     1   0     7000.    ! Beam energy  
'KBE2'     1   0     2212     ! 11: electron, 22: photon 2212: proton
'IRE2'     1   0        1     ! 0:  beam 3 has no structure 
*                             ! 1:  beam 2 has structure 
'NFLA'     1   0        4     ! (D=5) nr of flavours used in str.fct
* +++++++++++++++ Hard subprocess selection ++++++++++++++++++
'IPRO'     1   0        2     ! (D=1) 
*                             !  2: J/psi g
*                             !  3: chi_c
'I23S'      1    0       0    ! (D=0) select 2S or 3S state
'IPOL'      1    0       0    ! (D=0)  VM->ll (polarization study)
'IHFL'      1    0       4    ! (D=4) produced flavour for IPRO=11
*                             ! 4: charm
*                             ! 5: bottom				        
'PTCU'      1    0       1.   ! (D=0) p_t **2 cut for process
* ++++++++++++ Parton shower and fragmentation ++++++++++++
'NFRA'     1     0       1    ! (D=1) Fragmentation on=1 off=0
'IFPS'     1     0       3    ! (D=3) Parton shower
*                             ! 0: off
*                             ! 1: initial state PS
*                             ! 2: final state PS
*                             ! 3: initial and final state PS
'IFIN'     1     0       1    ! (D=1) scale switch for FPS
*                             ! 1: 2(m^2_1t+m^2_2t)    
*                             ! 2: shat     
*                             ! 3: 2(m^2_1+m^2_2)     
'SCAF'     1     0       1.   ! (D=1) scale factor for FPS
'ITIM'     1     0       0    ! 0: timelike partons may not shower
*                             ! 1: timelike partons may shower
'ICCF'     1     0       1    ! (D=1) Evolution equation 
*                             ! 0: DGLAP
*                             ! 1: CCFM
*                             ! 2: PB TMD evolution
* +++++++++++++ Structure functions and scales +++++++++++++
'IRAM'     1     0       0    ! (D=0) Running of alpha_em(Q2)
*                             ! 0: fixed
*                             ! 1: running
'IRAS'     1     0       1    ! (D=1) Running of alpha_s(MU2)
*                             ! 0: fixed alpha_s=0.3 
*                             ! 1: running
'IQ2S'     1     0       3    ! (D=1) Scale MU2 of alpha_s
*                             !  1: MU2= 4*m**2 (only for heavy quarks)
*                             !  2: MU2 = shat(only for heavy quarks)
*                             !  3: MU2= 4*m**2 + pt**2
*                             !  4: MU2 = Q2
*                             !  5: MU2 = Q2 + pt**2
*                             !  6: MU2 = k_t**2
'SCAL'     1    0     1.0     !  scale factor for renormalisation scale
'SCAF'     1    0     1.0     !  scale factor for factorisation scale*
*'IGLU'    1    0     1201    ! (D=1010)Unintegrated gluon density 
*                             ! > 10000 use TMDlib (i.e. 101201 for JH-2013-set1)
*                             !  1201: CCFM set JH-2013-set1 (1201 - 1213)
*                             !  1301: CCFM set JH-2013-set2 (1301 - 1313)
*                             !  1001: CCFM J2003 set 1 
*                             !  1002: CCFM J2003 set 2 
*                             !  1003: CCFM J2003 set 3 
*                             !  1010: CCFM set A0
*                             !  1011: CCFM set A0+
*                             !  1012: CCFM set A0-
*                             !  1013: CCFM set A1
*                             !  1020: CCFM set B0
*                             !  1021: CCFM set B0+
*                             !  1022: CCFM set B0-
*                             !  1023: CCFM set B1
*                             !  1: CCFM old set JS2001
*                             !  2: derivative of collinear gluon (GRV)
*                             !  3: Bluemlein
*                             !  4: KMS
*                             !  5: GBW (saturation model)
*                             !  6: KMR
*                             !  7: Ryskin,Shabelski
* ++++++++++++ BASES/SPRING Integration procedure ++++++++++++
'NCAL'     1    0    50000    ! (D=20000) Nr of calls per iteration for bases
'ACC1'     1    0     1.0     ! (D=1) relative prec.(%) for grid optimisation
'ACC2'     1    0     0.5     ! (0.5) relative prec.(%) for integration
* ++++++++++++++++++++++++++++++++++++++++++++++++++++++++++++
*'INTE'    1    0      0      ! Interaction type (D=0)
*                             ! = 0 electromagnetic interaction
*'KT1 '    1    0      0.44   ! (D=0.0) intrinsic kt for beam 1
*'KT2 '    1    0      0.44   ! (D=0.0) intrinsic kt for beam 2
*'KTRE'    1    0      0.35   ! (D=0.35) primordial kt when non-trivial 
*                             ! target remnant is split into two particles
* Les Houches Accord Interface
'ILHA'     1    0      0      ! (D=10) Les Houches Accord  
*                             ! = 0  use internal CASCADE
*                             ! = 1  write event file
*                             ! = 10 call PYTHIA for final state PS and remnant frag
* path for updf files
* 'UPDF'   './share'
\end{verbatim}
\end{footnotesize}

\section{Program Installation}
\cascade\ now follows the standard AUTOMAKE
convention. To install the program, do the following
\begin{footnotesize}
\begin{verbatim}
1) Get the source from http://www.desy.de/~jung/cascade

tar xvfz cascade-XXXX.tar.gz
cd cascade-XXXX

2) Generate the Makefiles (do not use shared libraries)
./configure --disable-shared --prefix=install-path --with-lhapdf="lhapdflib_path" 
--with-tmdlib="TMDlib_path"  --with-hepmc="hepmc_path"

with (as example):
lhapdflib_path=/Users/jung/MCgenerators/lhapdf/6.2.1/local
TMDlib_path=/Users/jung/jung/cvs/TMDlib/TMDlib2/local
hepmc_path/Users/jung/MCgenerators/hepmc/HepMC-2.06.09/local
3) Compile the binary
make

4) Install the executable and PDF files
make install 

4) The executable is in bin
run it with:
export CASEED=1242425
export HEPMCOUT=outfile.hepmc

cd $install-path/bin

./cascade < $install-path/share/cascade/LHE/steering-DY-MCatNLO.txt

\end{verbatim}
\end{footnotesize}
 
\section*{Acknowledgments.}
FG acknowledges the support and hospitality of DESY, Hamburg, where part of this work started.  
FH acknowledges the hospitality and support of DESY, Hamburg and of CERN, Theory Division while parts of this work were being done. 
SB, ALi and MM are grateful the DESY Directorate for the support in the framework of Cooperation Agreement between MSU and DESY 
on phenomenology of the LHC processes and TMD parton densities.
MM was supported by a grant of the foundation for the advancement of theoretical physics and mathematics ``Basis" 20-1-3-11-1.
STM thanks the Humboldt Foundation for the Georg Forster research fellowship  and 
gratefully acknowledges support from IPM.    
ALe acknowledges funding by Research Foundation-Flanders (FWO) (application number: 1272421N).
QW and HY acknowledge the support by the Ministry of Science and Technology under grant No. 2018YFA040390 and 
by the National Natural Science Foundation of China under grant No. 11661141008. 

 \newpage
 \section{Program Summary}
 
{\em Title of Program:} \cascade\ 3.1.0 \\ \\
{\em Computer for which the program is designed and others on which it is
operable:}   any with standard Fortran 77 (gfortran)\\ \\
{\em Programming Language used:}  FORTRAN 77 \\ \\
{\em High-speed storage required:}  No \\ \\
{\em Separate documentation available: } No \\ \\
{\em Keywords: } QCD, TMD parton distributions.\\ \\
\\ \\
{\em Method of solution:}  
Since measurements involve complex cuts and multi-particle final states, the 
ideal tool for any theoretical description of the data is a Monte Carlo 
event generator which generates initial state parton showers according to Transverse Momentum Dependent (TMD) parton densities, in a backward evolution, which  follows the evolution equation as used for the determination of the TMD. \\ \\
{\em Restrictions on the complexity of the problem:}  
Any LHE file (with on-shell or off-shell) initial state partons can be processed.\\ \\
{\em Other Program used:}  \PYTHIA\ ({\it version $>$ 6.4}) for final state parton shower and hadronization,  {\sc Bases/Spring}  5.1 
for integration (both supplied with the program package), \\
\tmdlib\ as a library for TMD parton densities. \\ \\
{\em Download of the program:} \verb+http://www.desy.de/~jung/cascade+\\ \\
{\em Unusual features of the program:}   None \\ \\
\newpage

\bibliographystyle{mystyle} 
\raggedright 

\providecommand{\etal}{et al.\xspace}
\providecommand{\href}[2]{#2}
\providecommand{\coll}{Coll.}
\catcode`\@=11
\def\@bibitem#1{%
\ifmc@bstsupport
  \mc@iftail{#1}%
    {;\newline\ignorespaces}%
    {\ifmc@first\else.\fi\orig@bibitem{#1}}
  \mc@firstfalse
\else
  \mc@iftail{#1}%
    {\ignorespaces}%
    {\orig@bibitem{#1}}%
\fi}%
\catcode`\@=12

\end{document}